\documentclass[fleqn,10pt]{wlscirep}
\usepackage[utf8]{inputenc}
\usepackage[T1]{fontenc}
\usepackage{graphicx}

\usepackage{soul}      
\usepackage{xcolor}    
\usepackage{soulutf8}  

\usepackage{caption}
\usepackage{subcaption}
\usepackage{adjustbox}
\usepackage{float}
\usepackage{wrapfig}

\usepackage{amsmath, amssymb}

\usepackage{algorithm}
\usepackage{algpseudocode}
\title{The Role of Fractal Dimension in Wireless Mesh Network Performance}

\author[1]{Marat Zaidyn}
\author[1]{Sayat Akhtanov}
\author[1*]{Dana Turlykozhayeva}
\author[1]{Symbat Temesheva}
\author[1]{Almat Akhmetali}
\author[1]{Alisher Skabylov}
\author[1]{Nurzhan Ussipov}

\affil[1]{Nonlinear Information Processes Laboratory (NIPL), Department of Electronics and Astrophysics, Al-Farabi Kazakh National University, 050040 Almaty, Kazakhstan}

\affil[*]{turlykozhayeva.dana@kaznu.kz}

\begin{abstract}
Wireless mesh networks (WMNs) depend on the spatial distribution of nodes, which directly influences connectivity, routing efficiency, and overall network performance. Conventional models typically assume uniform or random node placement, which inadequately represent the complex, hierarchical spatial patterns observed in practical deployments. In this study, we present a novel algorithm that constructs WMN topologies with tunable fractal dimensions, allowing precise control over spatial self-similarity. By systematically varying the fractal dimension, the algorithm generates network layouts spanning a continuum of spatial complexities, ranging from sparse fragmented clusters to dense, cohesive structures. Through NS-3 simulations, Key performance metrics including throughput, latency, jitter, and packet delivery ratio were evaluated across a range of fractal dimensions. Comparative evaluations against classical random, small-world, scale-free, grid and hierarchical tree networks models reveal that high-dimensional fractal topologies achieve enhanced resilience and throughput under equivalent conditions. These findings demonstrate the potential of fractal geometry as a design paradigm for scalable and efficient WMN architectures.

\end{abstract}

\begin{document}

\flushbottom
\maketitle
%
%
\thispagestyle{empty}

\section*{Introduction}

Wireless Mesh Networks (WMNs) have gained significant traction due to their flexibility, scalability, and self-organizing capabilities, enabling seamless communication in diverse applications, from home networks to urban and industrial deployments \cite{akyildiz2005wireless,kaur2023several}. Unlike traditional networks, WMNs use a distributed topology where nodes communicate directly without a central gateway, supporting multi-hop communication for extended coverage and dynamic reconfiguration \cite{benyamina2011wireless}. These characteristics have spurred extensive research on optimizing WMN topologies, with studies exploring routing protocols, node placement, energy efficiency, and fault tolerance \cite{pathak2010survey,tchinda2024energy,park2021proactive,ramphull2021review,caleb2023enhancing,turlykozhayeva2024experimental}.

The topology of a WMN significantly impacts performance metrics such as latency, reliability, and resilience \cite{herrera2023performance,cilfone2019wireless,barolli2015node,turlykozhayeva2024single}. Optimized topologies can reduce latency and congestion, improve fault tolerance through redundant paths, and enable scalability by integrating new nodes without degrading performance \cite{nurlan2021wireless, pasandideh2023systematic,muthaiah2008single,turlykozhayeva2024evaluating}. Traditional modeling approaches often use idealized spatial distributions, such as uniform grids or Poisson distributed nodes, which provide analytical tractability and baseline information on network behavior \cite{ding2023obstacle,mastan2024enhancing}. Complementing these, recent topology-aware network designs emphasize spatial structures that reflect the heterogeneous, clustered, or hierarchical arrangements observed in many practical deployments shaped by terrain, obstacles, and environmental factors \cite{zhang2022topology,wu2024topology,wen2021fractal,li2021topology}. 

Complex network theory offers a powerful framework for modeling systems with intricate connectivity patterns, such as social, biological, and technological networks. These systems are often characterized by properties such as small-world behavior, scale-free degree distributions, and self-similarity \cite{song2005self,molontay2015fractal,perera2017network}. In particular, self-similarity, where structures exhibit recurring patterns on multiple scales, is typically quantified by the fractal dimension, a measure of spatial or topological complexity derived from the scaling behavior of covering algorithms \cite{fronczak2024scaling,kovacs2021comparative,zhang2007self,lepek2025beyond}. Networks exhibiting finite fractal dimensions differ fundamentally from compact, hub-dominated topologies, as they lack densely interconnected cores and exhibit slower decay in box-counting curves \cite{rosenberg2018survey}. These characteristics naturally align with the topological constraints of wireless systems, where spatial organization directly shapes connectivity and communication efficiency. In this context, fractal networks provide a valuable theoretical foundation for investigating the role of topological structure in communication performance. A wide range of network covering algorithms have been developed to estimate the fractal dimensions of complex networks, including greedy coloring methods and degree-prioritized searches \cite{song2007boxcovering,wei2013box,zhang2014fuzzy}, as well as more recent strategies incorporating multi-objective optimization, stochastic sampling, and centrality-based heuristics \cite{giudicianni2021community,rosenberg2017minimal,schneider2012box}. These tools have enabled the quantification of both global self-similarity and local node-level fractal behavior \cite{cohen2004fractal,silva2012local}, which in turn have supported applications such as resilience analysis, vulnerability detection, and node influence ranking \cite{li2011dimension,ramirez2020fractional}.

However, it is worth noting that the impact of varying fractal dimensions on the performance of WMNs remains largely underexplored in the existing literature. While previous studies have examined the structural properties of complex networks, particularly in contexts such as resilience and influence analysis, few have systematically investigated how fractal dimension influences communication metrics such as throughput, latency, jitter, and delivery ratio in WMNs.

In this article, we propose an algorithm in which node distributions in WMNs are explicitly constructed according to fractal-based structures. Rather than inferring the dimensionality from empirical networks, this approach enables direct control over the fractal dimension, transforming it from a passive descriptive parameter into an active design variable. This framework is especially relevant for spatial systems, where node placement governs not only connectivity and signal coverage but also routing paths, interference dynamics, and overall network performance. In WMNs, the layout of the nodes is a primary determinant of the system behavior, particularly in practical environments that deviate from uniform or random configurations. Realistic deployments, shaped by terrain, access constraints, or human planning, often exhibit spatial heterogeneity on multiple scales. Incorporating fractality into network design thus offers a principled method for investigating the interplay between spatial complexity and performance, with the potential to inform more robust, scalable, and efficient wireless architectures.

\section*{Results}

\begin{figure}[ht]
\centering
\includegraphics[width=0.72\textwidth]{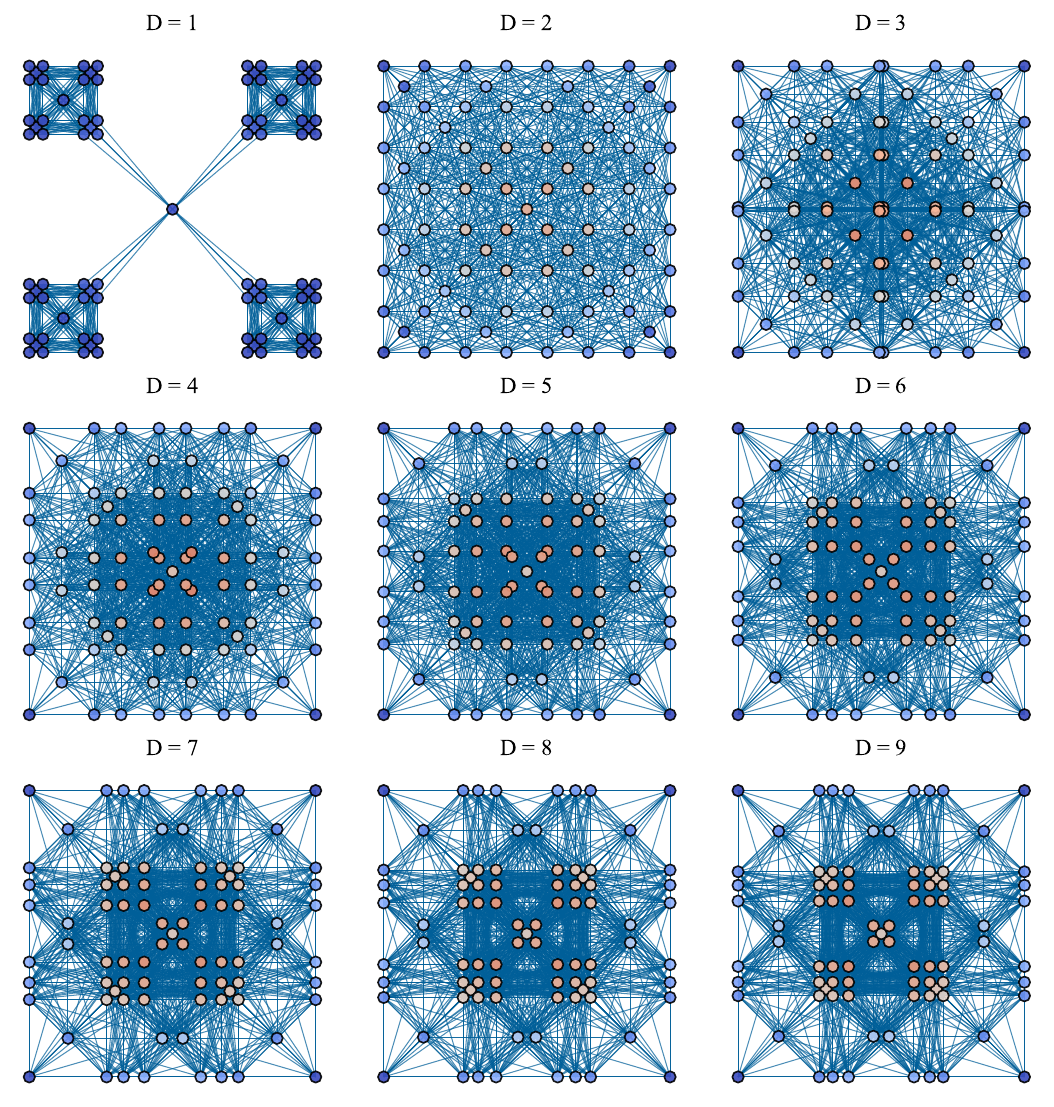}
\caption{Network topologies generated with fractal dimensions from  \( D = 1 \) to  \( D = 9 \). 
Low-dimensional networks exhibit fragmented, sparse structures, whereas higher-dimensional embeddings result in denser and more uniformly connected graphs. 
Node color indicates the average link capacity per node,  reflecting its local connectivity.}
\label{fig:topology}
\end{figure}

To investigate the relationship between fractal dimensionality and network performance, a series of WMNs with dimensions ranging from \( D \in [1, 9] \) were constructed using the proposed algorithm and evaluated through NS-3 simulations under consistent protocol and traffic conditions. Figure~\ref{fig:topology} presents the spatial topologies of the generated networks.

Figure~\ref{fig:topology} illustrates that at a low fractal dimension (\( D = 1 \)), the network fragments into sparsely connected components, forming isolated clusters with weak interconnections. This fragmentation results from the inherent spatial sparsity caused by the recursive process of low-dimensional generation. As the fractal dimension increases, the network becomes increasingly integrated and spatially uniform. At \( D = 2 \), the topology resembles a grid-like tiling, while at \( D = 3 \), more cross-linked central paths emerge, forming distinctive cruciform structures. Interestingly, from \( D \in [4, 8] \), the generated networks converge toward a stable topological regime characterized by dense, isotropic interconnectivity and robust multi-hop pathways. This suggests a structural saturation effect, where further increases in fractal dimension yield diminishing changes in spatial arrangement. When \( D = 7 \) and continuing through \( D = 9 \), ring-shaped or circular substructures begin to appear within the network topology. These formations represent the emergence of locally clustered subnetworks embedded within the overall topology, rather than a uniformly dispersed node distribution. This shift suggests that higher fractal dimensions promote localized aggregation, likely because of recursive spatial overlap during the point-generation process.

\begin{figure}[ht]
    \centering
    \includegraphics[width=1\linewidth]{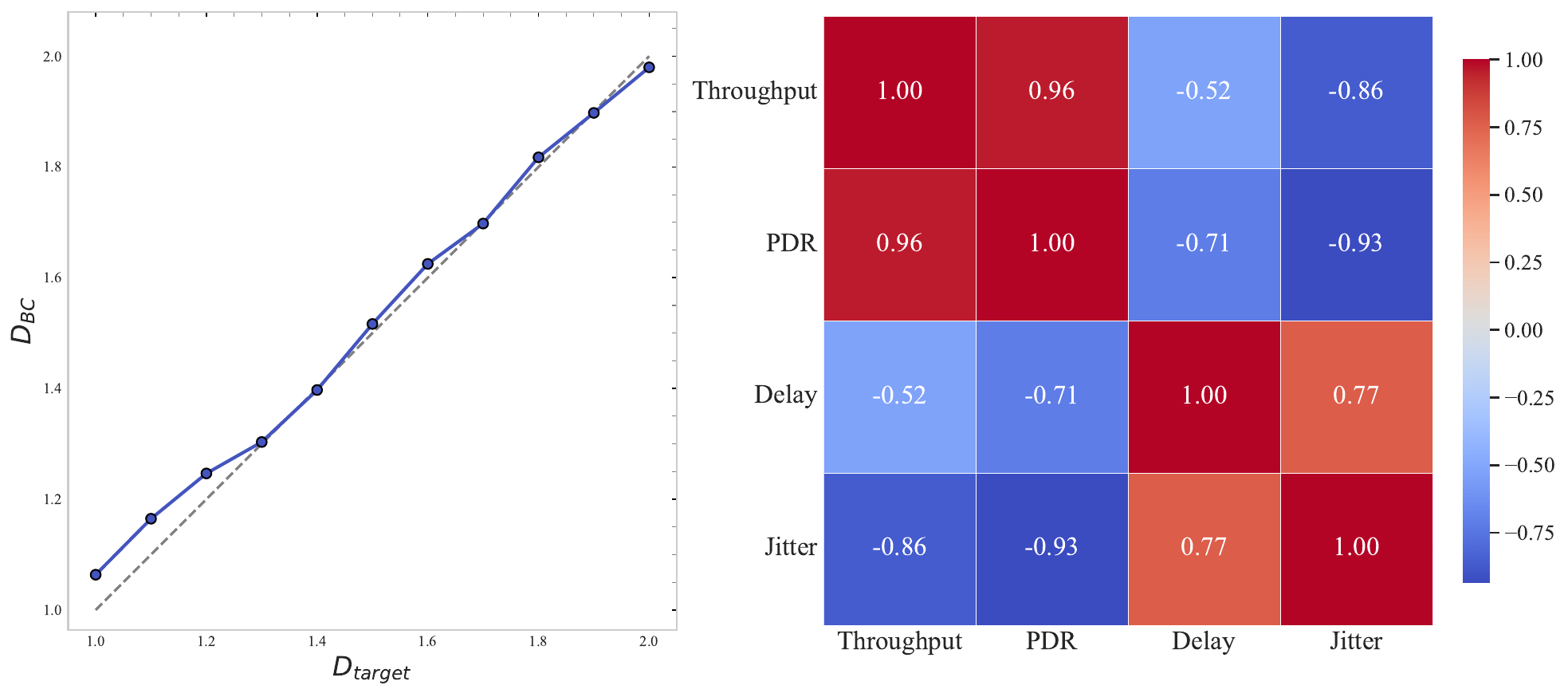}
\caption{
Fractal dimension validation and dependence of performance metrics. 
Left: Box-counting estimation of fractal dimension across spatial configurations with \( N = 10^5 \) nodes. The solid line(\( D_{\text{BC}}  \)),  denotes measured values, while the dotted line (\( D_{\text{target}}  \)) marks the ideal identity line, confirming the accuracy of the embedding algorithm. 
Right: Pearson correlation matrix among performance metrics. Throughput and PDR exhibit moderate positive correlation, whereas delay and jitter are largely uncorrelated with other metrics, indicating low redundancy across evaluation dimensions.
}
    \label{fig:boxcountconeff}
\end{figure}

\begin{figure}[ht]
        \centering
        \includegraphics[width=1\linewidth]{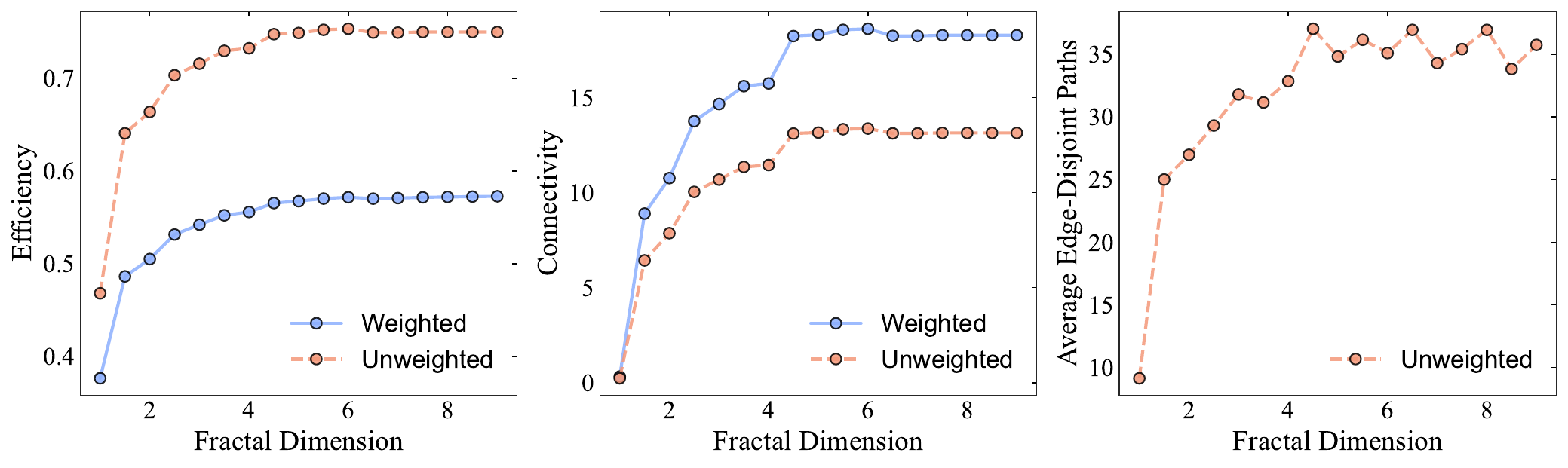}
        \caption{
        Global efficiency, algebraic connectivity, and average edge-disjoint paths (EDP) versus fractal dimension \( D = 1 \) to \( D = 10 \). Unweighted networks show higher efficiency but lower connectivity, and EDP is shown for unweighted networks only
        }
        \label{fig:fractal_main}
\end{figure}

To assess the accuracy of the proposed algorithm, the box-counting method was applied to a high resolution configuration with \( N = 10^5 \) nodes (Figure~\ref{fig:boxcountconeff}). The estimated box-counting fractal dimensions show a linear relationship with the target fractal dimension, confirming the algorithm’s ability to generate a spatial node distribution with controllable and verifiable fractal properties. The slight deviations observed at the edge of the curve are attributed to the inherent limitations of the box-counting technique, such as sparse cell occupancy at large scales and resolution-dependent estimation error. These results confirm the accuracy and dimensional controllability of the proposed algorithm in generating accurate node distributions. 
\begin{figure}[ht]
\centering
\includegraphics[width=1\textwidth]{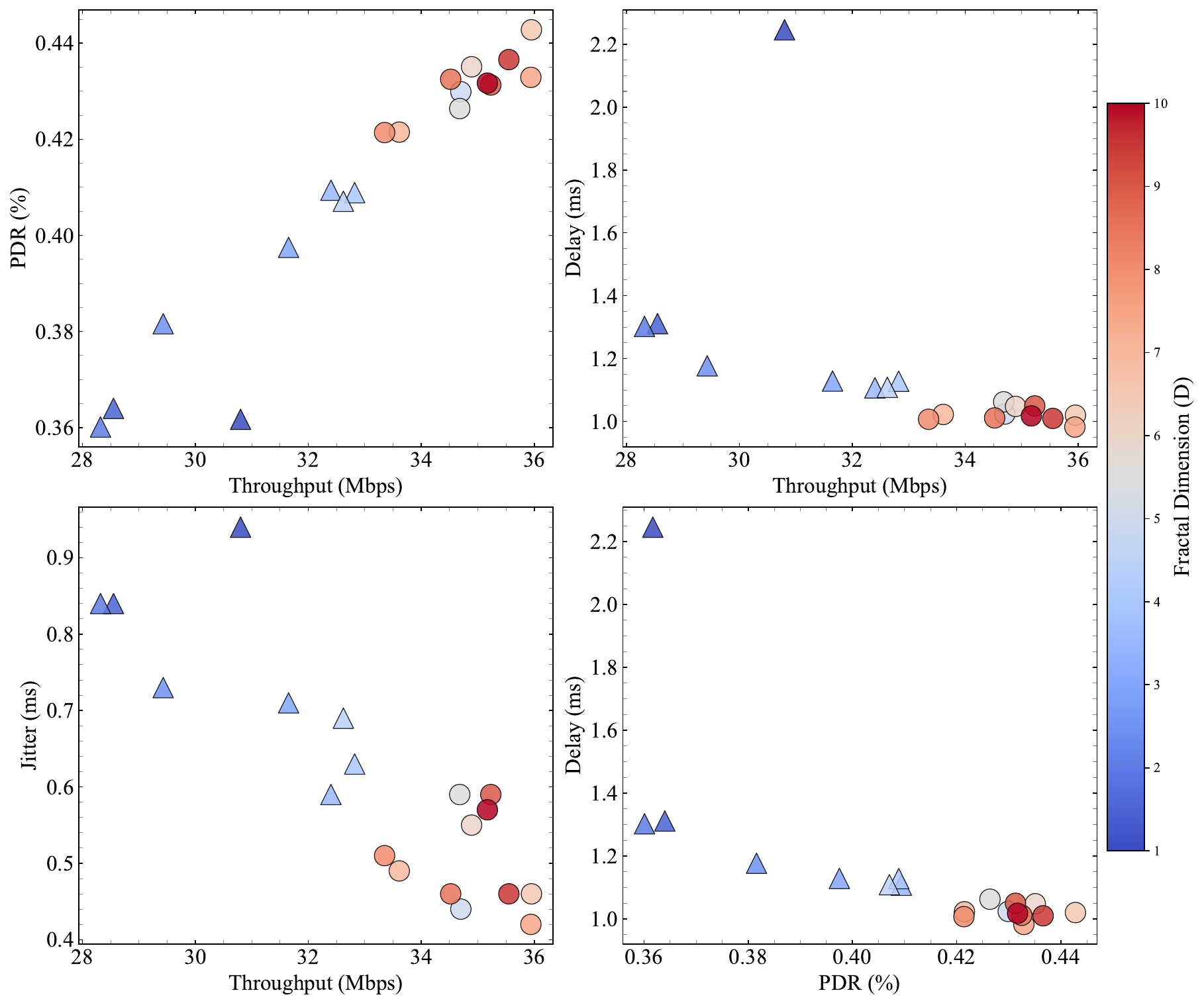}
\caption{Pairwise relationships between key performance metrics across fractal dimensions. Each point represents a network configuration; marker shapes indicate low ($D < 5$, triangles) and high ($D\geq5$, circles) fractal dimensions.Higher-dimensional networks consistently show better performance across all metric combinations.}
\label{fig:metrics}
\end{figure}

To test the correlation among performance metrics, pairwise Pearson correlation coefficients were computed (Figure~\ref{fig:boxcountconeff}). It is revealed that throughput and PDR are moderately correlated, reflecting their shared dependence on effective data transmission. Other metrics pairings, including delay versus jitter and delay versus PDR, exhibit weak correlations. These weak correlations show that each metric reflects a different aspect of performance. This explains why certain graphs (e.g., throughput versus delay or PDR versus delay) in Figure~\ref{fig:metrics} were not plotted, as they provide limited additional information due to weak mutual predictability.

Figure~\ref{fig:metrics} presents the joint distribution of key performance metrics such as throughput, PDR, jitter, and end-to-end delay over the full range of fractal dimensions. Networks with \( D \geq 5 \) consistently occupy regions associated with superior performance, forming compact clusters characterized by high throughput, high PDR, and low latency and jitter. In contrast, lower-dimensional networks (\( D < 5 \)) appear more scattered and exhibit substantially degraded performance. In particular, the network at \( D = 1 \) exhibits poor performance. Both throughput and PDR exhibit notable declines, while delay and jitter increase, showing poor communication. This reflects the fragmented topology (Figure~\ref{fig:topology}) characterized by poor linking and disjoint clusters.

\begin{figure}[ht]
        \centering
        \includegraphics[width=1 \linewidth]{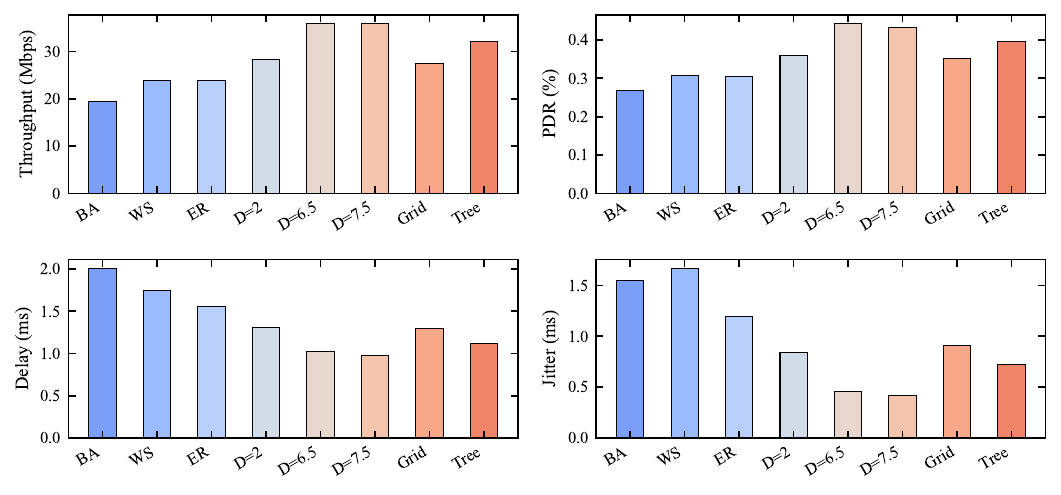}
        \caption{
        Comparative performance analysis of fractal and baseline networks. 
        Quality-of-service (QoS) metrics—including throughput, packet delivery ratio (PDR), end-to-end delay, and jitter are evaluated across fractal, scale-free (BA), small-world (WS), random (ER), grid and  hierarchical tree topologies under the same simulation conditions.
        }
        \label{fig:baseline_main}
\end{figure}

To reveal the structural origins of these performance trends, topological indicators including global efficiency, algebraic connectivity, and average edge-disjoint paths were further examined (Figure~\ref{fig:fractal_main}). Global efficiency is defined as the average of the inverse shortest path lengths between all node pairs. At low fractal dimensions, \( D = 1 \), the network is highly fragmented with low efficiency. For  \( D =2 \) and \( D =3 \) , connectivity improves and the clusters become more spatially distributed, leading to a clear increase in efficiency, although the topology is still not fully stabilized. Around \( D = 4.5 \), a topological transition point is reached: the network becomes spatially isotropic, and subsequent increases in \( D \) lead to structurally similar but denser layouts, the network approaches a mature, cohesive topology, and efficiency plateaus, reflecting structural saturation. Interestingly, comparing weighted and unweighted networks reveals that the unweighted efficiency is consistently higher with a similar trend. This occurs because the weighted network penalizes long-range or weak links with lower weights, effectively increasing the “cost” of certain paths. In contrast, unweighted networks treat all links equally, allowing more alternative paths to contribute fully to efficiency, thereby producing higher average values.

\begin{figure}[ht]
        \centering
        \includegraphics[width=1\linewidth]{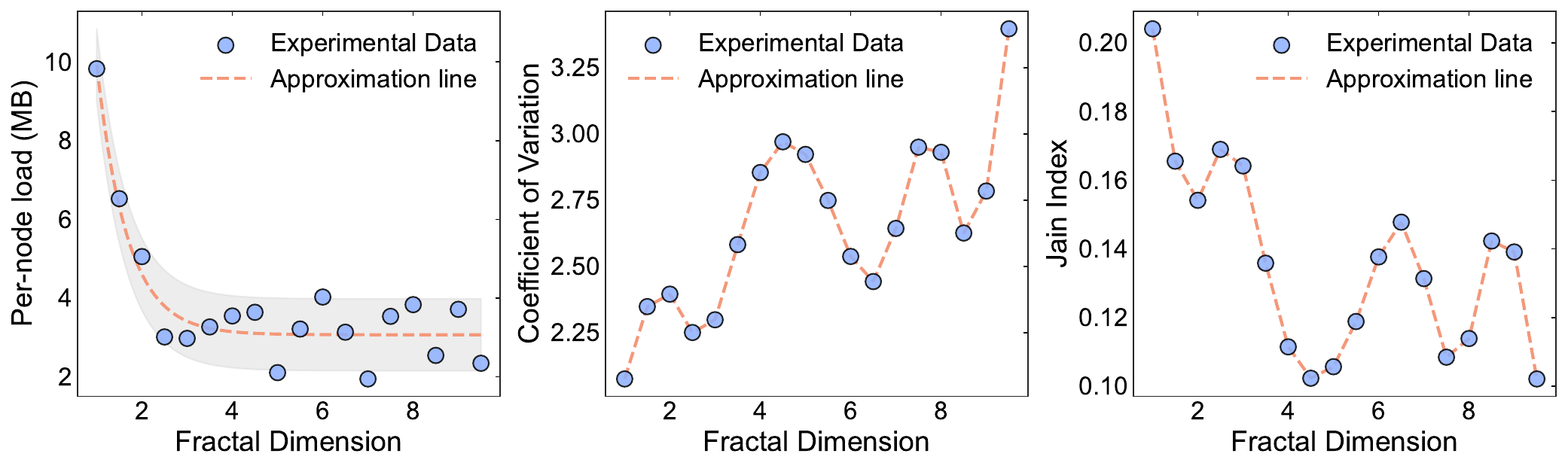}
        \caption{
        Per-node load, coefficient of variation (CV), and Jain’s fairness index versus fractal dimension $D$. 
Mean load decreases with $D$, while CV and Jain’s index capture non-monotonic shifts in load balancing.
        }
        \label{fig:pernode}
\end{figure}

Algebraic connectivity, representing spectral robustness and derived from the Laplacian matrix weighted by link capacity, increases steadily between \( D = 1 \) and  \( D = 4.5 \), reflecting improved interconnectivity and reduced path fragmentation. Beyond \( D = 4.5 \), the growth plateaus, indicating structural saturation where further increases in \( D \) yield marginal connectivity gains.  A similar trend is observed for the unweighted network, though overall connectivity is lower because unweighted links ignore capacity differences, reducing the effective contribution of weak or long-range connections. This transition coincides with the stabilization of global efficiency and confirms that early topological shifts are the primary drivers of performance enhancement. To further quantify path diversity, we calculated the average number of edge-disjoint paths between node pairs in the unweighted network. EDPs grow up to \(D = 4.5\), reflecting increasing redundancy and alternative routing options, then plateau with small fluctuations due to the discrete nature of path counting. By Menger’s theorem, the  EDPs between nodes is bounded by the minimum cut (or degree)~\cite{menger1927}, aligning with the observed saturation.

To evaluate load balancing beyond per-node averages, we first examined the total received bytes per node. While the per-node load alone shows a general trend of decreasing variance with increasing fractal dimension, a more detailed view using the coefficient of variation (CV) and Jain’s fairness index reveals complex fluctuations across \(D\). As shown in Figure~\ref{fig:pernode}, the CV rises from \(D=1\) to \(D=2\), slightly decreases, increases up to \(D=4.5\), then drops toward \(D=7\), rises to \(D=8\), briefly decreases, and increases again toward \(D=9\). In contrast, Jain’s index follows an opposite but similarly non-monotonic pattern. Notably, in the intermediate range \(D=5.5\)-\(7.5\), the node load becomes relatively uniform and path diversity is high, resulting in peak network performance consistent with the throughput, latency, jitter, and packet delivery metrics as Figure~\ref{fig:metrics}. These observations highlight that while per-node averages provide a baseline, the combination of load distribution and path diversity drives optimal performance in specific fractal dimension ranges.

To further contextualize the advantages of high-dimensional fractal topologies, networks were compared with several classical baselines including random geometric graphs, scale-free, small-world, grid and hierarchical tree configurations under equivalent traffic conditions (Figure~\ref{fig:baseline_main}). For this comparison, three representative fractal networks with dimensions \( D = 2 \), \( D = 6.5 \), and \( D = 7.5 \) were selected. When \( D = 2 \), the quasi-regular structure resembles grid-like arrangement. \( D = 6.5 \) and \( D = 7.5 \) were selected as representative high-performance configurations, based on the performance peak observed across all QoS metrics. These dimensions correspond to structural saturation points, beyond which further increases in D yield diminishing returns in performance. Including these values allowed us to benchmark near-optimal fractal networks against classical topologies. The results confirm that fractal-based networks, particularly those at higher \( D \), consistently perform better than baseline networks. In particular, the performance of \( D = 2 \)  is nearly identical to that of the grid topology, indicating that low-dimensional fractal layouts can emulate deterministic spatial structures. While BA networks exhibit degree-based hierarchy and topological self-similarity, they lack the spatial self-similarity of our fractal networks. BA topologies form centralized hubs, whereas fractal networks display recursive spatial clustering, enabling better path diversity and load distribution. As a result, fractal networks consistently outperform BA in key performance metrics. However, hierarchical structures like BA still offer competitive performance, though slightly inferior due to their hub-centric bottlenecks under load. These findings demonstrate that a higher fractal dimension yields a dual benefit: improved structural cohesion and enhanced performance.

\section*{Discussion}

The results indicate that adjusting the fractal dimension of the network topology can effectively influence both structural and performance characteristics in WMNs. As the \( D \) increases, key metrics such as throughput, delay, and delivery ratio exhibit significant improvement. These findings underscore the potential of fractal dimension as an active design parameter in the structuring of WMNs. Rather than serving solely as a descriptive metric, controlled variation of fractal dimension enables the systematic shaping of both topological and functional characteristics. 

While these findings offer promising insights, certain methodological considerations merit further exploration. The current generation pipeline, which maintains a fixed node count of 
\( N = 85 \) across all fractal dimensions, enables controlled comparisons but can limit the generalizability of the results, particularly with respect to scalability under varying network sizes. In practical deployments, WMNs often span hundreds or thousands of nodes and their performance is shaped by emerging phenomena such as routing bottlenecks, radio interference, and power asymmetries. Future work should aim to extend the fractal generation algorithm to support flexible node counts while preserving the statistical self-similarity of the spatial layout. A promising direction involves the use of recursive subdivision schemes with tunable depth and stochastic variability, enabling the creation of dense or sparse regions while maintaining a globally coherent fractal structure. 

In addition, a constant communication radius was adopted across all simulations. While this choice ensures comparability, it inherently ties the spatial scale of connectivity to a fixed node density, which may obscure the nuanced interplay between local clustering, signal coverage, and interference dynamics. At higher fractal dimensions, nodes are positioned closer, forming more cohesive internal structures that can also enhance path diversity in routing. Future work could explore adjusting the communication radius dynamically based on local node density or estimated fractal dimension to achieve more accurate performance evaluations and better guide real-world transmission strategies.

While the structural coherence of the algorithm was verified through box-counting analysis, deviations from power-law scaling were observed at both low and high-dimensional extremes, particularly in configurations with small node counts. These findings indicate that although the generation method demonstrates near-linear scaling in larger networks, it produces quasi-fractal structures when \( N \) is small, where the statistical basis for dimension estimation becomes unstable due to sparse spatial coverage. This quasi-fractal behavior is consistent with real-world constraints, where ideal self-similarity is uncommon and resource limitations necessitate coarse approximations. For future work, more rigorous validation techniques, such as modular box-covering strategies, should be investigated to better quantify network structure under finite-size effects.

From a theoretical standpoint, WMN performance is governed by fractal geometry, connectivity and routing efficiency. 
For a node set of fractal dimension $D$, the mass–radius law $N(s)\propto s^{D}$ implies that with fixed range $r$, the mean degree scales as $\langle k\rangle\!\propto\! r^{D}$~\cite{grassberger1983,Falconer2003,penrose2003}. 
Larger $\langle k\rangle$ increases clustering and reach, creating denser communities and redundant links. 
By Menger’s theorem, the number of edge-disjoint paths between two nodes is determined by the minimum cut, so higher degree yields more alternative routes~\cite{Diestel2005}. 
Spectral theory complements this view: for the normalized Laplacian, algebraic connectivity $\lambda_2$ obeys
\begin{equation}
    \tfrac{1}{2}\phi(G)^2 \le \lambda_2 \le 2\phi(G),
\end{equation}
linking conductance $\phi(G)$ to resilience and expansion~\cite{Chung1997}. 
Global efficiency $E$ (Equation~\eqref{eq:efficiency}), the mean inverse shortest-path length, measures multi-hop effectiveness.  As $D$ increases, networks evolve from fragmented to efficient topologies.  In our simulations ($N=85$, $r=250$\,m), gains plateau for \(D \gtrsim 4\), due to a cut bound path diversity limited by minimum degree $\delta(G)$, and a geometric bound fixed $N$ and $r$ restricting diameter and $E$.  These bounds explain the saturation of $\lambda_2$, $E$, path diversity and fairness, while denser layouts still improve throughput, delivery ratio and delay.  Thus fractal dimension acts as a tuning parameter shaping connectivity, routing geometry and performance. The interplay between load distribution and path diversity explains why peak performance occurs around $D\approx 6.5$ rather than at higher dimensions. In this intermediate regime, connectivity and efficiency are already high, while traffic remains well-balanced across nodes, as reflected in low CV and high Jain’s index (Figure~\ref{fig:pernode}).  At larger $D$ ($\geq 8$), new links primarily add redundant neighbors without creating independent routes: the cut bound limits and the geometric bound limits efficiency. As a result, flows concentrate on highly connected nodes, raising CV and lowering fairness, which constrains throughput and latency improvements despite higher density.  This explains the alignment betweenFigures~\ref{fig:fractal_main}and~\ref{fig:pernode}, and shows that achieve near-optimal network performance emerges when structural connectivity and balanced load coincide.

Looking ahead, the use of the fractal dimension to guide the network structure opens new avenues for research in wireless systems. Beyond WMNs, systems such as sensor swarms, drone fleets, and even urban vehicular networks can benefit from topological configurations that are more scalable and adaptive. Moreover, integrating fractal design principles into emerging paradigms such as graph neural networks (GNNs), programmable radio networks, or multi-agent reinforcement learning could enable the co-optimization of structure and behavior in distributed systems. 

\section*{Method}

This section presents an algorithm that constructs WMN topologies by explicitly setting the fractal dimension \( D \) as an input parameter to control the spatial distribution of the nodes. The nodes are positioned according to the specified fractal dimension, and the edges of the network are formed by connecting pairs of nodes within a fixed communication radius \( r \) . The resulting fractal-based topologies are then analyzed using NS-3 simulations, where key performance indicators such as throughput, end-to-end delay, jitter, and PDR are measured under uniform traffic scenarios. To implement the proposed algorithm, the following steps are performed:

\begin{enumerate}
    \item Construct fractal-based topology according to a certain fractal dimension   \( D \)  ;
    \item Build the network by connecting node pairs within a certain radius  \( r \).
\end{enumerate}

\subsubsection*{Fractal-based Topology Construction}

WMNs can be formally described as undirected graphs \( G(N, E) \), where \( N = \{1, 2, \dots, n\} \) denotes the set of nodes (e.g. routers or devices) and \( E = \{1, 2, \dots, m\} \) denotes the set of communication links. The network structure is described by an adjacency matrix \( A \in \{0, 1\}^{|N| \times |N|} \), defined as:

\begin{equation}
a_{ij} = 
\begin{cases} 
1, & \text{if node } i \text{ is directly connected to node } j, \\
0, & \text{otherwise.}
\end{cases}
\end{equation}

To generate topologies with controllable spatial self-similarity, a recursive coordinate generation algorithm was developed, drawing inspiration from fractal geometry and operating within a normalized two-dimensional space. In this algorithm, each node recursively produces four child nodes arranged along quadrant directions, with internode spacing generated by a fractal dimension parameter \( D \). The produced topology forms a structured, irregular spatial distribution that reflects key characteristics of fractal patterns. The full procedure is given in Algorithm~\ref{alg:fractal}.

\begin{figure}[htbp]
    \centering
    \begin{minipage}{0.8\textwidth}
        \centering
        \begin{algorithm}[H]
        \caption{Fractal Coordinate Generation}\label{alg:fractal}
        \begin{algorithmic}[1]
            \State \textbf{Input:} Target number of nodes $N$, fractal dimension $D$
            \State \textbf{Output:} First $N$ node coordinates $(x_i, y_i)$
            \State Initialize empty list: $\mathit{points} \gets \emptyset$
            \State Compute maximum recursion depth: $\mathit{depth}_{\max} \gets \left\lfloor \frac{\log(3N)}{\log 4} \right\rfloor$
            \State Set initial center and scale: $(x_0, y_0) \gets (0.5, 0.5)$, $s_0 \gets 0.5$
            \Function{Generate}{$x, y, s, d$}
                \If{$d \leq 0$}
                    \State \Return
                \EndIf
                \State Append $(x, y)$ to $\mathit{points}$
                \State Update scale: $s' \gets s \cdot 4^{-1/D}$
                \For{\textbf{each} $(\Delta x, \Delta y) \in \{(\pm1, \pm1)\}$}
                    \State \Call{Generate}{$x + \Delta x \cdot s', y + \Delta y \cdot s', s', d-1$}
                \EndFor
            \EndFunction
            \State \Call{Generate}{$x_0, y_0, s_0, \mathit{depth}_{\max}$}
            \State \Return First $N$ points from $\mathit{points}$
        \end{algorithmic}
        \end{algorithm}
    \end{minipage}
\end{figure}

The scaling behavior highlights the mathematical definition of the fractal dimension, originally introduced by Hausdorff \cite{hausdorff1918dimension} and popularized by Mandelbrot \cite{mandelbrot1983fractal}. The fractal dimension \( D \) is defined as:

\begin{equation}
D = \frac{\ln N_0}{\ln b},
\label{eq:fractal_dimension}
\end{equation}

where \( N_0 \) is the number of self-similar segments generated in each recursive step and \( b \) is the spatial scaling ratio. From equation~\eqref{eq:fractal_dimension}, the expression can be rearranged as:

\begin{equation}
b = N_0^{1/D}.
\end{equation}

The initial number of parent nodes is fixed (\( N_0 = 4 \)), forming a quadrant fractal structure at each iteration.
At recursion depth \( k+1 \), the spatial distance is accordingly updated by:

\begin{equation}
s_{k+1} = s_k \times 4^{-1/D}.
\end{equation}

The total number of nodes \( N \) generated up to depth \( k \) follows a geometric series:

\begin{equation}
N = \sum_{i=0}^{k} 4^i = \frac{4^{k+1} - 1}{3}.
\label{eq:fixed_nodes}
\end{equation}

Inverting this relation gives an explicit formula for determining the required recursion depth for a desired network size :

\begin{equation}
\text{depth} = \frac{\log(3N + 1)}{\log 4} - 1.
\label{eq:max_depth}
\end{equation}

\begin{figure}[htbp]
\centering
\includegraphics[width=1\textwidth]{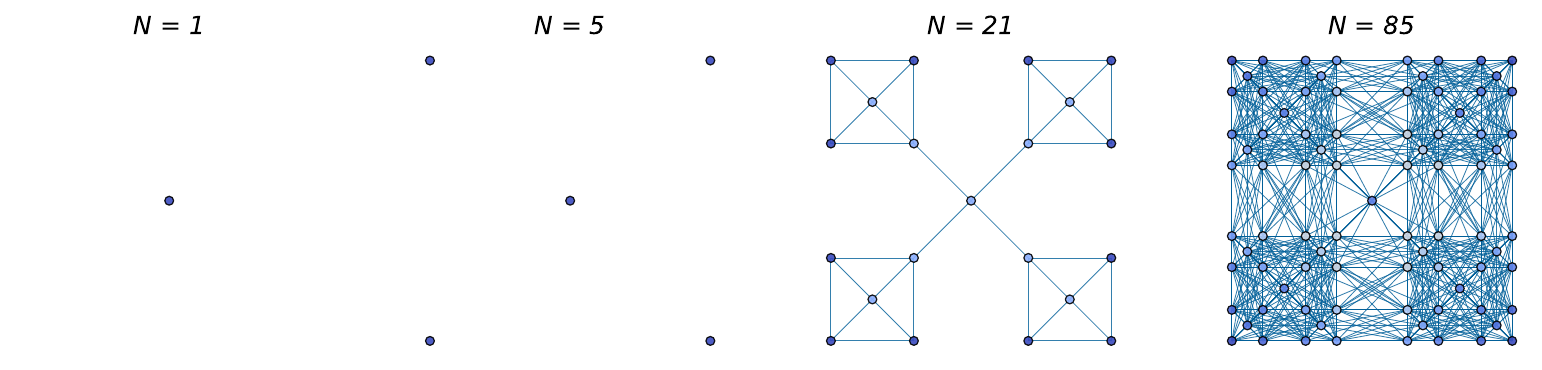}
\caption{Example of recursive generation of network topology with \( D = 1.6 \). The reduced quadrants display repeating subpatterns, reflecting the self-similarity typical of fractal geometry.}
\label{fig:stream}
\end{figure}

\begin{table}[htbp]
    \centering
    \caption{Number of nodes depending on recursion depth, each recursion level increases the number of nodes.}
    \label{tab:depth_vs_nodes}
    \begin{tabular}{cc}
        \hline
        Depth & Nodes \\
        \hline
        0 & 1 \\
        1 & 5 \\
        2 & 21 \\
        3 & 85 \\
        4 & 341 \\
        5 & 1365 \\
        \vdots & \vdots \\
        \hline
    \end{tabular}
\end{table}

Equation~\eqref{eq:max_depth} describes how the number of nodes grows with recursion depth, increasing by a factor of four at each level and resulting in exponential growth, as shown in Table~\ref{tab:depth_vs_nodes}. Figure~\ref{fig:stream} illustrates an example implementation of the proposed algorithm.

To establish a rigorous baseline for evaluating the structural and communication advantages of fractal-based networks, three classical generative network models were constructed: Erdős–Rényi (ER), Watts–Strogatz (WS), and Barabási–Albert (BA), all generated under consistent conditions of network size and mean degree. Specifically, ER networks were generated using the $G(n, p)$ model with $n = 85$ nodes and edge probability $p = \frac{k}{n-1}$, targeting a mean degree $k = 4.5$. WS networks were initialized as regular ring lattices with each node connected to its $k = 4$ nearest neighbors and then rewired with a probability of $p = 0.1$ to introduce small-world properties. BA networks were created using a preferential attachment mechanism, starting with a small fully connected seed and sequentially adding nodes, each establishing $m = 2$ links based on the degree of the existing node. To ensure a valid comparison with the fractal networks, all generated networks were required to be fully connected, with repeated sampling applied as needed. Random seeds were fixed to maintain reproducibility and consistent layout configurations. In addition to the classical ER, WS, and BA networks, we introduced two deterministic topologies to broaden the structural comparison: a regular grid network and a hierarchical tree network. The grid network was constructed as a $9 \times 9$ two-dimensional lattice ($81$ nodes), augmented by $4$ additional nodes placed at the centers of the four corner cells, resulting in a total of $85$ nodes.The grid layout provides a spatially uniform and deterministic structure with regular local neighborhoods, emulating idealized deployments in structured environments such as factory floors or urban grids. The hierarchical network was generated as a balanced tree with branching factor $r = 4$ and height $h = 3$, yielding exactly $85$ nodes. Each non-leaf node connects to $4$ child nodes in a recursive structure, forming a layered topology that emphasizes centralized hierarchy and minimal path diversity.

\begin{figure}[htbp]
\centering
\includegraphics[width=0.8\textwidth]{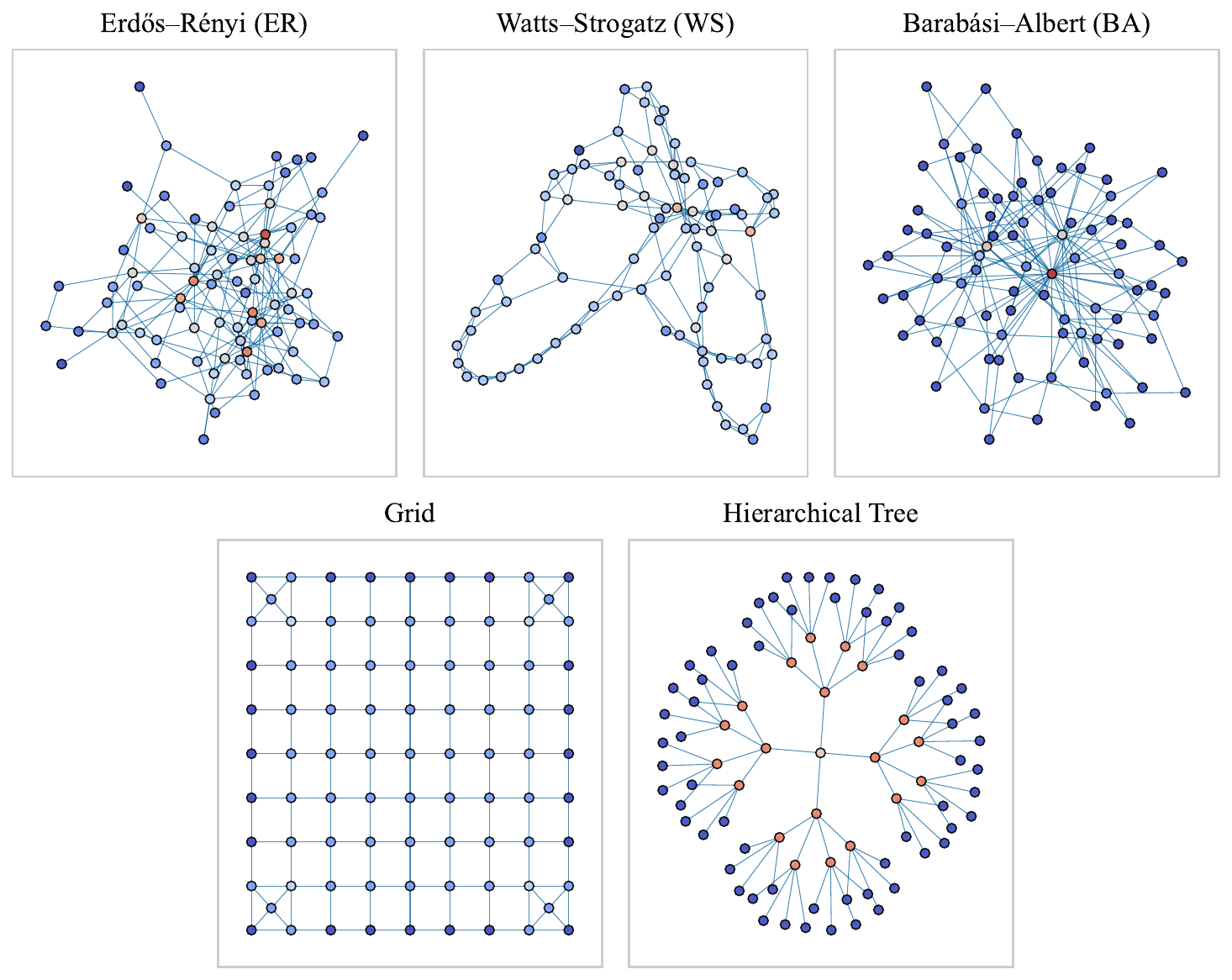}
\caption{Visualization of classical network models (ER, WS, BA,grid and hierarchical tree) used as baseline topologies for comparison. All configurations have identical node counts and mean degrees to ensure fair evaluation. Node colors represent average link capacity, highlighting local connectivity patterns.}
\label{fig:networksc}
\end{figure}

These baseline models represent, respectively, random (ER), small-world (WS), scale-free (BA), spatially regular (grid), and hierarchically modular (tree) regimes commonly studied in the network topology literature. They serve as essential structural archetypes for evaluating the topological distinctiveness and impact on performance of high-dimensional fractal geometries. The corresponding topologies are illustrated in Figure~\ref{fig:networksc}. This diversity provides a set of contrastive regimes for benchmarking the behavior of fractal-based network designs.

\subsubsection*{NS-3 Based Evaluation of Fractal WMNs}

To evaluate the influence of fractal dimension on WMN performance, simulations were conducted in NS-3, measuring key metrics such as throughput, end-to-end delay, jitter, and PDR across a range of fractal dimensions. The simulation parameters are summarized in Table~\ref{tab:sim_params}.

\begin{table}[htbp]
\centering
\caption{NS-3 Simulation Parameters}
\label{tab:sim_params}
\begin{tabular}{ll}
\hline
 \textbf{Parameter} & \textbf{Value} \\
\hline
version & 3.40 \\
 Number of Nodes & 85 \\
 Simulation Area & 500\,m $\times$ 500\,m \\
 Connectivity Radius&250 m\\
 Wireless Standard & IEEE 802.11ac \\
 Propagation Model & Log-Distance Propagation Loss \\
 Antenna Model & Omnidirectional \\
 Node Mobility & Static \\
 Routing Protocol & OLSR (with static fallback) \\
 Traffic Model & UDP (OnOffApplication) \\
 Traffic Rate & 100\,Mbps \\
 Number of Flows & 100 random source–destination pairs \\
 Transport Protocol & UDP \\
 Packet Size & 1024\,Bytes \\
 Application Start Time & Uniform[0\,s, 10\,s] \\
 Simulation Duration & 50\,s \\
 Energy Model & Li-ion, 100\,J initial energy \\
 TX/RX Current & 17.4\,mA / 19.7\,mA \\
 Performance Metrics & Throughput, Delay, Jitter, Packet Loss \\
\hline
\end{tabular}
\end{table}

The simulation environment consisted of 85 static nodes deployed using the proposed spatially heterogeneous, self-similar fractal layout across a \(500\,\mathrm{m} \times 500\,\mathrm{m}\) area. The node positions were generated using the proposed algorithm (Algorithm~\ref{alg:fractal}), with the fractal dimension \( D \in [1, 10] \). The resulting self-similar and spatial distributions were visually verified.

All nodes were provisioned with IEEE 802.11ac compliant wireless interfaces to support high-throughput communication.Wireless signal propagation was modeled using the Log-Distance Propagation Loss Model, while timing characteristics were captured using the Constant Speed Propagation Delay Model. Omnidirectional antennas were used, and the communication radius was configured to ensure reliable multi-hop connectivity throughout the network.

Routing was performed using the Optimized Link State Routing (OLSR) protocol, integrated through the \textit{Ipv4ListRoutingHelper} to enable compatibility with static routing mechanisms. Network traffic was generated using 100 independent UDP flows between randomly selected source-destination pairs. Each flow transmitted 100,000 packets of 1024 bytes at a constant rate of 100\,Mbps using the \textit{OnOffApplication} model. The start times were uniformly distributed within the initial 10 seconds of the simulation to avoid synchronization artifacts.

Network performance data was collected using the built-in \textit{FlowMonitor} module, with custom Python scripts used to parse the trace files and compute aggregate metrics. Specifically, throughput was calculated as the total number of successfully received bits per unit of time; end-to-end delay was defined as the average time a packet took to travel from source to destination, accounting for queuing and propagation delays; jitter was quantified as the mean deviation of interpacket arrival times; PDR was computed as the fraction of packets dropped over the total sent. Each simulation scenario was repeated 30 times to ensure statistical robustness and all results are reported as mean values with 95\% confidence intervals.

In addition to the standard metrics, including throughput, end-to-end delay, jitter, and packet delivery ratio (PDR) obtained via NS-3's FlowMonitor module, we evaluated additional topological and load-balancing indicators to quantify path diversity and traffic distribution.

The global efficiency of the network is defined as the average of the inverse shortest path lengths between all node pairs:
\begin{equation}
E = \frac{1}{N(N-1)} \sum_{i \neq j} \frac{1}{d_{ij}},
\label{eq:efficiency}
\end{equation}
where $d_{ij}$ denotes the shortest path length between nodes $i$ and $j$, considering weighted or unweighted graphs as specified.

Algebraic connectivity $\lambda_2$ is computed as the second-smallest eigenvalue of the graph Laplacian matrix
\begin{equation}
L = D - A,
\end{equation}
where $D$ is the degree matrix and $A$ is the adjacency matrix, weighted by link capacity for the weighted case.

The average number of edge-disjoint paths (EDPs) between node pairs is obtained using a max-flow algorithm (e.g., Ford-Fulkerson) on the unweighted graph and then averaged over all pairs.

To assessload balancing, we analyzed the total received bytes per node. The coefficient of variation (CV) for node loads $x = (x_1, \dots, x_N)$ is defined as
\begin{equation}
\mathrm{CV} = \frac{\sigma}{\mu},
\end{equation}
where $\sigma$ and $\mu$ represent the standard deviation and mean of the node loads, respectively.

Jain's fairness index $J$ is defined as~\cite{jain1984}:
\begin{equation}
J = \frac{\left( \sum_{i=1}^{N} x_i \right)^2}{N \sum_{i=1}^{N} x_i^2},
\end{equation}
quantifying the fairness of resource allocation, where $J = 1$ corresponds to perfect fairness and $J = 1/N$ indicates the worst case. Note that CV and $J$ are related by
\begin{equation}
J = \frac{1}{1 + \mathrm{CV}^2},
\end{equation}
explaining their inverse fluctuations.

\section*{Data availability}

The simulation data generated and analysed during this study are available at the following GitHub repository: \href{https://github.com/Zaidyn-marat/Fractal-based-WMNs-Topology}{https://github.com/Zaidyn-marat/Fractal-based-WMNs-Topology}. The repository includes the full fractal topology generator, NS-3 simulation scripts, and performance evaluation code used to produce the results presented in this paper. Additional datasets are available from the corresponding author upon reasonable request.

\bibliography{sample}
\section*{Acknowledgements}

We would like to express our sincerest gratitude to the
Al-Farabi Kazakh National University for supporting this
work by providing computing resources (Department of
Physics and Technology). 

\section*{Funding}

This research was funded by the Committee of Science of the Ministry of Science and Higher Education of the Republic of Kazakhstan (Grant No. AP19674715).

\section*{Author contributions statement}

N.U. and M.Z. conceived and designed the study. N.U. developed the methodology. M.Z. and S.A. implemented the software. M.Z. and A.A. conducted validation. S.A. performed formal analysis and data curation. N.U. led the investigation. S.A. provided resources. S.T. and M.Z. prepared the original manuscript draft. A.S. contributed to manuscript refinement and provided feedback on the writing. D.T. contributed to manuscript review and editing. M.Z. created the visualizations. N.U. supervised the project. All authors have read and approved the final manuscript.

\end{document}